\pdfoutput=1
\documentclass{optica-article}

\articletype{Research Article}

\usepackage{graphicx}
\usepackage{bm}
\usepackage{float}
\usepackage{caption}
\usepackage{subcaption}
\usepackage{amsmath}
\usepackage[latin1]{inputenc}
\usepackage{mathtools}
\usepackage{gensymb}
\usepackage{upgreek}
\usepackage{orcidlink}
\usepackage{xcolor}

\usepackage{lineno}

\begin{document}
\title{Lasing at a Stationary Inflection Point: erratum}

\author{
A.~Herrero-Parareda~\orcidlink{https://orcid.org/0000-0002-8501-5775}\authormark{1}, 
N.~Furman~\orcidlink{https://orcid.org/0000-0001-7896-2929}\authormark{1}, 
T.~Mealy~\orcidlink{https://orcid.org/0000-0001-7071-9705}\authormark{1}, 
R.~Gibson~\orcidlink{https://orcid.org/0000-0002-2567-6707}\authormark{2}, 
R.~Bedford~\orcidlink{https://orcid.org/0000-0002-0457-0081}\authormark{3}, 
I.~Vitebskiy~\orcidlink{https://orcid.org/0000-0001-8375-2088}\authormark{2}, and 
F.~Capolino~\orcidlink{https://orcid.org/0000-0003-0758-6182}\authormark{1,*}
}

\address{\authormark{1}Department of Electrical Engineering and Computer Science, University of California, Irvine, CA 92617, USA\\
\authormark{2}Air Force Research Laboratory, Sensors Directorate, Wright-Patterson Air Force Base, Ohio 45433, USA \\
\authormark{3}Air Force Research Laboratory, Materials and Manufacturing Directorate, Wright-Patterson Air Force Base, Ohio 45433, USA}
\email{\authormark{*}f.capolino@uci.edu} 



\begin{abstract*}
This erratum provides an updated fitting function for the lasing threshold of finite-length cavities operating at a stationary inflection point (SIP) or regular band edge (RBE) resonance, clarifying their asymptotic scaling with the number of unit cells of the periodic cavity.
\end{abstract*}

\vspace{1cm}

In this erratum, we present a revised fitting function for the magnitude of the SIP lasing threshold $n_{S,th}^{\prime\prime}(N)$ of a finite-length SIP-ASOW cavity with a length of $N$ unit cells. Despite the fitting provided in Ref.~\cite{parareda_lasing_2023} looking reasonable to the eye, it did not show the asymptotic scaling law $n_{S,th}^{\prime\prime}(N)\sim N^{-3}$ for large $N$. We provide here a better fitting to show such asymptotic trend. The affected figures by the imprecise fitting we provided in the original manuscript Ref.~\cite{parareda_lasing_2023} were Fig. 4 and Fig. 7b. Those figures should be replaced with the first two figures provided here. We also add a third figure to show the error function between the fitting functions $n_{S,fit}^{\prime\prime}(N)$ (in the original manuscript and the revised one presented here) and the threshold function $n_{S,th}^{\prime\prime}(N)$ to highlight the asymptotic behavior for large $N$. In the original manuscript, this threshold was fitted as $n_{S,th}^{\prime\prime}(N) = -0.016 \times N^{-3} + 7 \times 10^{-8}$ for $N \in [5,40]$, where $n^{\prime\prime}<0$ denotes gain. However, this form does not vanish as $N \to \infty$, and thus fails to rigorously reflect the expected asymptotic scaling. Here, we provide an updated fitting function for the threshold magnitude and show that it actually provides a good fitting for large $N$:
\begin{equation}
n_{S,fit}^{\prime\prime}(N) =- \frac{1}{eN^3 + fN},
\label{eq:rightFitting}
\end{equation}
where $e = 50.3485$ and $f= 3.96 \times 10^3$, fitted over $N \in [30,40]$. For $N > 9$, the $1/(e N^3)$ term dominates with respect to $1/(fN)$, capturing the expected $N^{-3}$ asymptotic scaling law.

For the RBE lasing threshold $n_{R,th}^{\prime\prime}$, the same fitting form is used $n_{R,fit}^{\prime\prime}(N) = -(gN^3 + hN)^{-1}$, with parameters $g = 37.54$ and $h = 8.044 \times 10^3$, also fitted over $N \in [30,40]$. Here, the $1/(gN^3)$ term becomes dominant for $N > 15$.

\begin{figure}[H]
    \centering\includegraphics[width=0.7\linewidth]{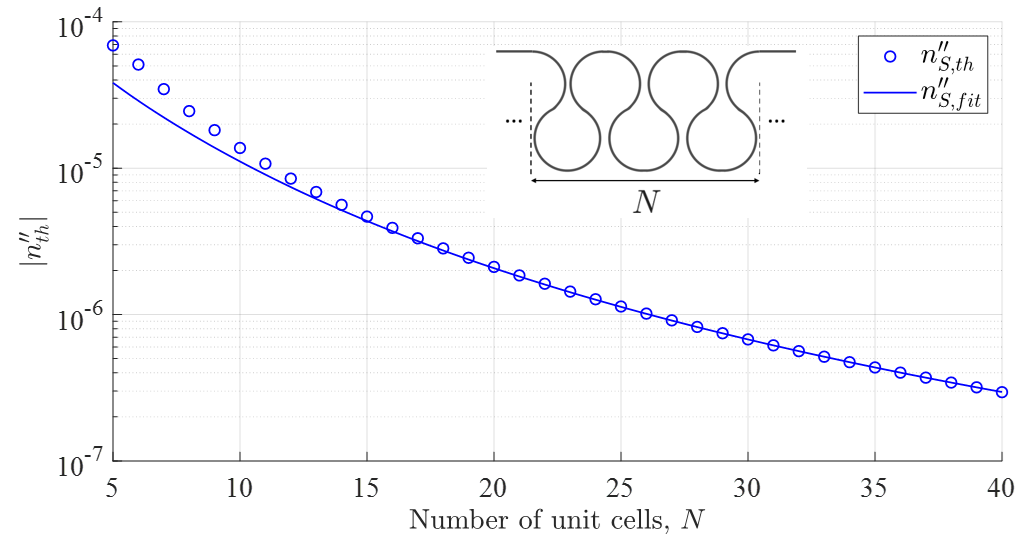}
    \caption{Same plot as in Fig. 4 in the original manuscript \cite{parareda_lasing_2023}, but with the proper fitting function for the SIP lasing threshold, showing the $1/N^3$ asymptotic scaling for the SIP lasing threshold.}
    \label{fig:fitting1}
\end{figure}

\begin{figure}[H]
    \centering  \includegraphics[width=0.6\linewidth]{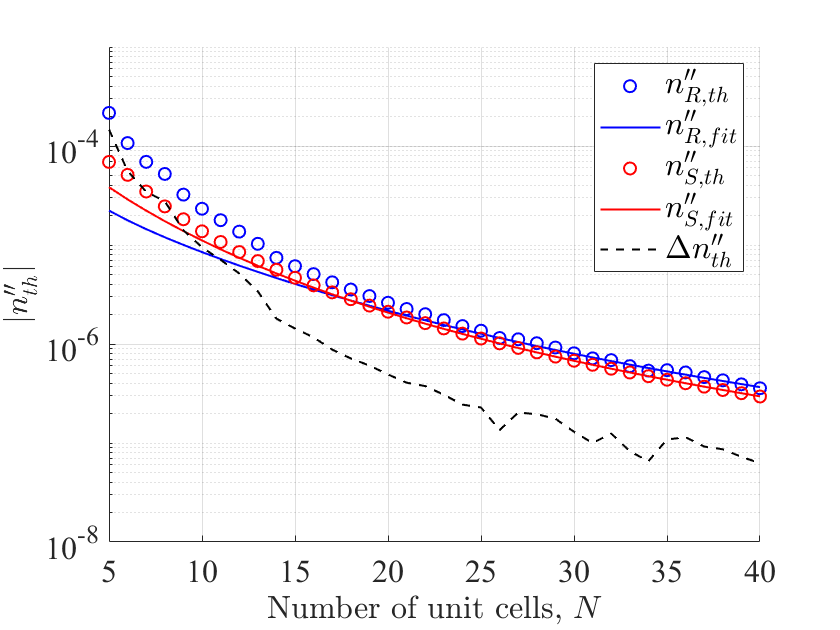}
    \caption{
    Same plot as in Fig. 7b in the original manuscript \cite{parareda_lasing_2023}, but with the proper fitting functions for the SIP and RBE, showing the $1/N^3$ asymptotic scaling. The black dashed line shows the magnitude of the threshold difference, $\Delta n_{th}^{\prime\prime} = n_{R,th}^{\prime\prime} - n_{S,th}^{\prime\prime}$.
    }
    \label{fig:fittings}
\end{figure}

The updated fitting for the magnitude of the SIP lasing thresholds $n_{S,th}^{\prime\prime}$ is shown in Figure~\ref{fig:fitting1}, which replaces Fig. 4 in the original manuscript. The updated comparison between the  fittings for the SIP and the RBE thresholds is shown in Fig.~\ref{fig:fittings} (replacing Fig.7b of the original manuscript), where the magnitude of the SIP and RBE thresholds are shown as red and blue dots. The solid lines of the same color represent their respective fittings introduced above. Besides observing that the magnitude of both SIP and RBE lasing thresholds asymptotically decrease as $1/N^3$, we also observe that $|n_{R,th}^{\prime\prime}|$ is always larger than $|n_{S,th}^{\prime\prime}|$.

\begin{figure}[H]
    \centering
    \includegraphics[width=0.6\linewidth]{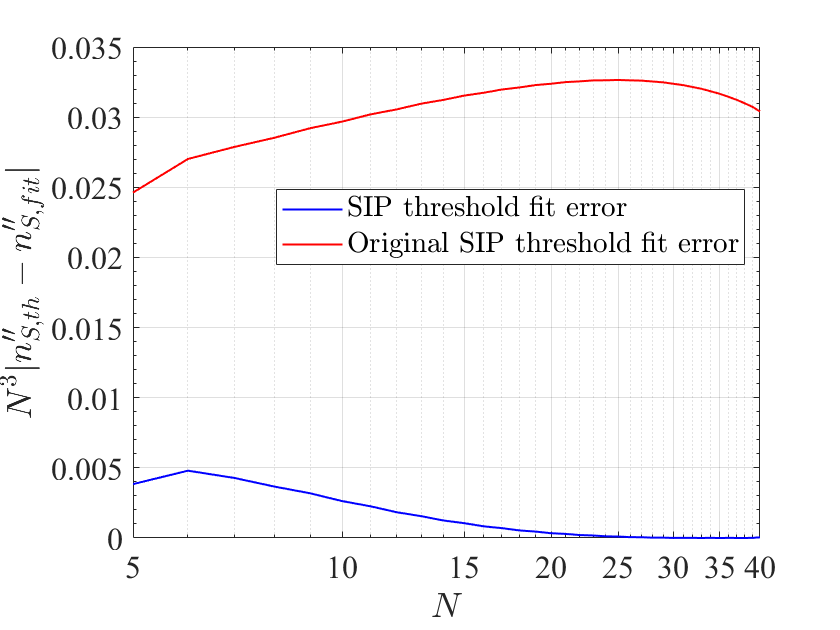}
    \caption{
    Error of the SIP lasing fitting function for large $N$, showing the asymptotic $1/N^3$ trend. The error tends to a small value even when multiplied by $N^3$ to emphasize the trend at large $N$.
    }
    \label{fig:fittingErrors}
\end{figure}

To assess the better quality of this fitting of the SIP lasing threshold with respect to the one in Ref.~\cite{parareda_lasing_2023}, Figure~\ref{fig:fittingErrors} plots the weighted fitting errors of the magnitude of the SIP lasing thresholds, i.e., $N^3 \times |n_{S,th}^{\prime\prime} - n_{fit}^{\prime\prime}(N)|$, versus $N$. This scaling emphasizes how well the updated fitting function (in blue) approximates the SIP's dominant $N^{-3}$ trend at large $N$, especially compared to the fitting in the original manuscript (in red).

Indeed, assuming the generic asymptotic expansion of a function for large $N$

\[
\begin{aligned}
    n_{S,th}^{\prime\prime}(N) &=  - \left[ \frac{c_1}{N} + \frac{c_2}{N^2} + \frac{c_3}{N^3} + \frac{c_4}{N^4}  + \frac{c_5}{N^5}+ O(N^{-6}) \right],
\end{aligned}
\]

and using the provided fitting function in Eq. (\ref{eq:rightFitting}), 
the error multiplied by $N^3$ becomes

\[
N^3 \times |n_{S,th}^{\prime\prime} - n_{S,fit}^{\prime\prime}| = \left| c_1 N^2+ c_2 N + \left(c_3 - \frac{1}{e}\right) + \frac{c_4}{N} + \frac{c_5 + f/e^2}{N^2} + O(N^{-3}) \right|,
\]

In Fig.~\ref{fig:fittingErrors}, it is clear that the original fitting function in \cite{parareda_lasing_2023} did  not fully demonstrate the dominance of the $N^{-3}$ term, while the weighted error of the revised fitting function in (\ref{eq:rightFitting}) tends to a very low value  for large  $N$. From the fact that the weighted error does not grow, we conclude that $c_1=0$ and $c_2=0$. Furthermore, since the error of the revised fitting function for large $N$ is very small, the fitting $1/e$ value is a good representation of the $c_3$ coefficient of the threshold asymptotic expansion. Therefore, a flat $N^3$-weighted error in Figure~\ref{fig:fittingErrors} confirms that the dominant asymptotic behavior of the SIP threshold is indeed $\sim N^{-3}$.

In summary, the SIP and RBE lasing thresholds, and their difference, are approximated as  
 
\begin{equation}
\begin{split}
  n_{S,th}^{\prime\prime} &\approx - \frac{1}{e N^{3} + fN}, \\
  n_{R,th}^{\prime\prime} &\approx -  \frac{1}{g N^{3} + hN}, \\
  \Delta n_{th}^{\prime\prime} =  n_{R,th}^{\prime\prime}-  n_{S,th}^{\prime\prime} &\ \approx - \frac{e-g}{eg}\frac{1}{N^3} + \frac{he^2-fg^2}{e^2g^2}\frac{1}{N^5},
\end{split}
  \label{eq:SIPRBELasingThr}
\end{equation}

where the last one is obtained by applying the first-order Taylor expansion of the first two. The coefficients are $e = 50.3485$, $f = 3.96\times 10^{3}$, $g = 37.54$, and $h = 8.044\times 10^{3}$. These expressions replace Eq. (15) of the original manuscript \cite{parareda_lasing_2023}. The magnitude of the difference between the lasing thresholds associated with the RBE resonance and the SIP resonance, $\Delta n_{th}^{\prime\prime} = n_{R,th}^{\prime\prime} - n_{S,th}^{\prime\prime}$, is depicted as a dashed black line in Fig.~\ref{fig:fittings}. The difference  between the SIP and RBE lasing thresholds  scales asymptotically as $N^{-3}$, resulting in comparable thresholds for large waveguides, though in our results the SIP threshold is always lower than the RBE one, also verified by the expansion coefficients satisfying $e>g$.

\bibliography{LasingSIP}

@article{parareda_lasing_2023,
author = {A. Herrero-Parareda and N. Furman and T. Mealy and R. Gibson and R. Bedford and I. Vitebskiy and F. Capolino},
journal = {Opt. Mater. Express},
number = {5},
pages = {1290--1306},
publisher = {Optica Publishing Group},
title = {Lasing at a stationary inflection point},
volume = {13},
month = {May},
year = {2023},
url = {https://opg.optica.org/ome/abstract.cfm?URI=ome-13-5-1290},
doi = {10.1364/OME.481906},
}

\end{document}